\documentclass[12pt]{article}

\pdfoutput=1

\usepackage{color}
\usepackage{graphicx}
\usepackage{amsmath}
\usepackage{amssymb}
\usepackage{xspace}
\usepackage[small]{subfigure}
\usepackage[numbers,compress]{natbib}
\usepackage[hyperfootnotes=false]{hyperref}

\newlength{\fighskip} \fighskip=2pt
\newlength{\figvskip} \figvskip=3pt

\newcommand*{\figbox}[2]{{
  \def\figscale{#1}
  \def\arraystretch{0.8}
  \arraycolsep=0pt
  \begin{array}{c}
    \vbox{\vskip\figscale\figvskip
      \hbox{\hskip\figscale\fighskip
        \includegraphics[scale=\figscale]{#2}}}
  \end{array}}}
  
\usepackage{mciteplus} 
\usepackage{dcolumn}
\usepackage{bm}
\usepackage{verbatim}
\usepackage{amscd}
\usepackage{amsfonts}
\usepackage{setspace}
\usepackage{amsthm}
\usepackage{enumerate}
\usepackage{mathtools}

\newcommand{\blangle}{\bigl\langle}
\newcommand{\brangle}{\bigr\rangle}
\newcommand{\dlangle}{\langle\kern-1.5pt\langle}
\newcommand{\drangle}{\rangle\kern-1.5pt\rangle}
\newcommand{\bdlangle}{\blangle\kern-3pt\blangle}
\newcommand{\bdrangle}{\brangle\kern-3pt\brangle}

\newcommand*{\corr}[1]{\langle#1\rangle}
\newcommand*{\bcorr}[1]{\blangle#1\brangle}
\newcommand*{\ccorr}[1]{\dlangle#1\drangle}

\DeclareMathOperator{\Tr}{Tr}
\DeclareMathOperator{\OTOC}{OTOC}
\DeclareMathOperator{\LL}{L}
\DeclareMathOperator{\UU}{U}

\title{Efficient decoding for the Hayden-Preskill protocol}
\author{Beni Yoshida\\ 
{\normalsize\it Perimeter Institute for Theoretical Physics, Waterloo, Ontario N2L 2Y5, Canada} \\ \\
Alexei Kitaev\\ 
{\normalsize\it California Institute of Technology, Pasadena CA 91125, USA}}

\date{}

\begin{document}

\maketitle

\begin{abstract}
We present two particular decoding procedures for reconstructing a quantum state from the Hawking radiation in the Hayden-Preskill thought experiment. We work in an idealized setting and represent the black hole and its entangled partner by $n$ EPR pairs. The first procedure teleports the state thrown into the black hole to an outside observer by post-selecting on the condition that a sufficient number of EPR pairs remain undisturbed. The probability of this favorable event scales as $1/d_{A}^2$, where $d_A$ is the Hilbert space dimension for the input state. The second procedure is deterministic and combines the previous idea with Grover's search. The decoding complexity is $\mathcal{O}(d_{A}\mathcal{C})$ where $\mathcal{C}$ is the size of the quantum circuit implementing the unitary evolution operator $U$ of the black hole. As with the original (non-constructive) decoding scheme, our algorithms utilize scrambling, where the decay of out-of-time-order correlators (OTOCs) guarantees faithful state recovery. 
\end{abstract}



\section{Introduction}

Black holes have long fascinated theorists and presented them with various puzzles. Are different pieces of Hawking radiation completely uncorrelated or, rather, entangled on long time scales? Does information contained in infalling objects come out in the radiation and how can it be recovered? While our understanding of quantum gravity is still incomplete, such general questions can hopefully be answered using a very simple model. A black hole is characterized by a Hilbert space of dimension $d=2^S$, where the \emph{coarse-grained entropy}\footnote{In this paper, we define entropy with the binary logarithm.} $S$ is proportional to the horizon area. The quantum evolution over a sufficiently long time interval $t$ (longer than the so-called scrambling time) is described by a Haar-random  unitary operator $U$. This approach was pioneered by Page~\cite{Page93,Page93a}, who considered a black hole forming from particles in some pure state and partially evaporating. The emitted radiation $R$ and the remaining black hole $B$ are described by Hilbert spaces of dimensions $d_R$, $d_B$ such that $d_Rd_B=d$. The state $|\Psi\rangle$ of the whole system is pure but so complex that may be regarded as random. Page found that if $d_R\ll d_B$, then the entanglement entropy is with high accuracy equal to $\log_2d_R$. Thus, a chunk of radiation is uncorrelated unless it includes half of the original black hole. On the other hand, if $d_R\gg d_B$, then some correlations exist and their specific form depends on $|\Psi\rangle$.

\begin{figure}
\centering\(\displaystyle
\begin{array}{@{}c@{\hspace{2cm}}c@{}}
\figbox{1.0}{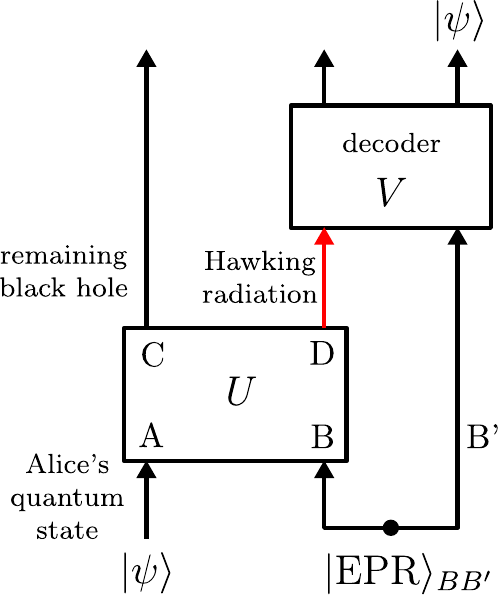} &
\figbox{1.0}{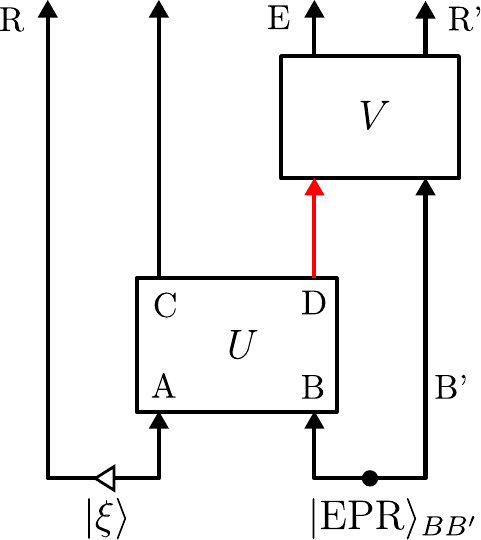}\: =\: \figbox{1.0}{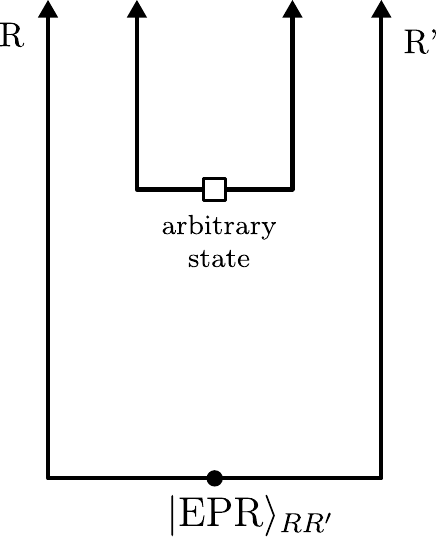}
\vspace{5pt}\\
\text{a)} & \text{b)}
\end{array}
\)
\caption{The Hayden-Preskill decoding problem (a) and its variant with a reference system (b).} 
\label{fig_Hayden_Preskill_decoding}
\end{figure}

Hayden and Preskill studied an interesting variant of the information recovery problem. Their paper~\cite{Hayden07} is both profound and fun to read! In short, Alice wants to destroy her confidential diary and tosses it into a black hole. One may model Alice's secret by a quantum state $|\psi\rangle$ of some system $A$ that is added to the original black hole $B$. Bob tries to spy on Alice by capturing some Hawking radiation (subsystem $D$ in Figure~\ref{fig_Hayden_Preskill_decoding}a). The important difference from Page's setting is that $B$ is maximally entangled with another system $B'$, which is in Bob's possession. In this situation, Bob does not have to wait until half of the black hole evaporates. In fact, $|\psi\rangle$ can be extracted from $D$ and $B'$ by applying some unitary decoder $V$, provided $\log_2d_D\ge\log_2d_A+\epsilon$. The parameter $\epsilon$ is related to the decoding fidelity and may be taken as a constant. Thus, the black hole acts as a quantum information mirror, reflecting whatever falls in it almost immediately. The delay is equal to the scrambling time $t_{\text{scr}}$ plus the time needed to radiate $\log_2d_A+\epsilon$ qubits. This ground-breaking work has led to recent studies that largely focused on the physics of scrambling~\cite{Sekino08, Lashkari13, Shenker:2013pqa, Maldacena:2016aa,Kitaev:2014t2}. It turns out that there are many good information scramblers, but black holes are the fastest ones among equilibrium systems at a given temperature: they satisfy the estimate $t_{\text{scr}}\approx(2\pi T)^{-1}\ln S$.

However, there is an important caveat. Hayden and Preskill showed that the decoding task is \emph{information-theoretically} possible in the sense that there exists a unitary operator $V$ that reconstructs $|\psi\rangle$ by acting only on the Hawking radiation $D$ and the auxiliary system $B'$. But it is not clear how complex this operator is and whether finding it from $U$ is a computationally tractable problem. One can argue that the decoding complexity is at least polynomial in $d_A$, i.e.\ exponential in the number of qubits that constitute Alice's diary. This can be seen from the classical analogue of the Hayden-Preskill problem, where Alice's secret $a$ and the black hole state $b$ are binary words, and $U$ is replaced by an invertible function. Bob having access to $B'$ means that he knows $b$. Let us consider $b$ as a fixed parameter and express the radiation state as $r=f(a)$. Since $r$ is just a small part of the overall state, the function $f$ need not be invertible; we may rather regard it as completely random. The condition $d_D\gg d_A$ guarantees that $a$ can be recovered from $r$, but the only general method to find $a$ is exhaustive search. 

Thus, the real question is how the decoding complexity scales with the black hole size. The answer is not obvious. On the one hand, Harlow and Hayden argued that it is exponentially hard to process the Hawking radiation of an old (more than half-evaporated) black hole so as to produce the standard EPR state~\cite{Harlow:2013tf}. We simply assume this state to be available. On the other hand, a physical process has recently been discovered that is akin to Hayden-Preskill decoding. First, Gao, Jafferis and Wall~\cite{Gao16} showed that the Einstein-Rosen bridge in the AdS black hole geometry can be made traversable if one arranges a momentary coupling (at time $0$) between the opposite boundaries so as to generate negative Casimir energy. In this setup, a signal sent from one boundary at a particular time $-t$ before the interaction is turned on reaches the other boundary at time $t$. Although the signal travels through the bulk, there is a holographically dual description strictly in terms of the boundaries. Its relation to the Hayden-Preskill problem and some aspects of the bulk-boundary correspondence were elucidated by Maldacena, Stanford and Yang~\cite{Traversable2017}. However, this particular decoding scheme works due to special properties of the operator $U$ at ``early times'', i.e.\ for $t<t_{\text{scr}}$.

Before attempting a solution for a random $U$ (or more generally, for the late times), let us slightly simplify the problem along the lines of the original Hayden and Preskill paper. Instead of estimating the worst-case recovery fidelity, we will take some average over $|\psi\rangle$. This idea is captured by a standard trick: one considers Alice's diary as part of a fixed entangled state $|\xi\rangle$ that also includes some reference system $R$. We assume that $|\xi\rangle=(I_R\otimes \Xi)|\text{EPR}\rangle_{RR'}$, where $R'$ represents the information content of Alice's diary and $\Xi:\,R'\to A$ is an isometric embedding. Bob's goal is to reconstruct $|\text{EPR}\rangle_{RR'}$ as shown in Figure~\ref{fig_Hayden_Preskill_decoding}b. The number $\log_2 d_A\ge\log_2d_R$ is interpreted as the increase of the coarse-grained black hole entropy. It follows from standard thermodynamics that $\ln d_A=E/T$, where $T$ is Hawking's temperature and $E$ is the energy of Alice's diary (including the rest energy).

\section{Notation, basic assumptions, and summary of results}

We will extensively use diagrams like those in Figure~\ref{fig_Hayden_Preskill_decoding}b. In general, nodes (e.g.\ $U$ and $|\xi\rangle$) are tensors, and the connecting lines represents the contraction of indices. A few additional rules formalize the idea that the upward direction is time. In particular, lines are labeled at places where they go vertically. The same line may carry a label $B$ at one point and $B'$ at a different point if it bends and reverses direction. Pairs of labels such as $B$ and $B'$ refer to dual Hilbert spaces. For each ket-vector $|\psi\rangle=\sum_{j}c_{j}|j\rangle\in A$, there is a dual vector $|\psi^*\rangle=\sum_{j}c_{j}^*|j\rangle\in A'$. It is just $\langle\psi|$ under a different name, but we keep them separate. Ket-vectors are associated with upward lines and bra-vectors with downward lines:
\begin{equation}
|\psi\rangle=\figbox{1.0}{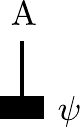}\,,\qquad\quad
|\psi^*\rangle=\figbox{1.0}{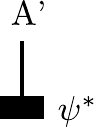}\,,\qquad\quad
\langle\psi|=\figbox{1.0}{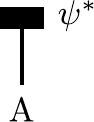}\,,\qquad\quad
\langle\psi^*|=\figbox{1.0}{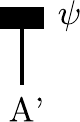}\,.
\end{equation}
For operators, we put their mathematical symbols in boxes and change $X$ to $X^T$ when the box is rotated $180^\circ$ degrees. For example,
\begin{equation}
\figbox{1.0}{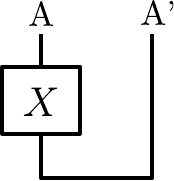} \,\,=\,\, \figbox{1.0}{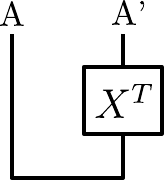}\,\,
= \sum_{j,k}X_{jk}|j,k\rangle.
\end{equation}
A dot on a line with label $B$ is equivalent to an overall factor of $d_B^{-1/2}$:
\begin{equation}
\figbox{1.0}{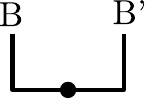}\: =|\text{EPR}\rangle_{BB'}
=\frac{1}{\sqrt{d_B}}\sum_{j}|j,j\rangle
=\: \frac{1}{\sqrt{d_B}}\,\figbox{1.0}{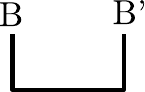}\:.
\end{equation}
A triangle corresponds to the embedding $\Xi:\,R'\to A$ multiplied by $d_R^{-1/2}$, but its exact meaning depends on the orientation. For example,
\begin{equation}
\figbox{1.0}{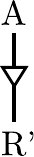}\: = \frac{1}{\sqrt{d_R}}\,\Xi\,,\qquad\quad
\figbox{1.0}{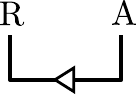}\: = (I_R\otimes \Xi)|\text{EPR}\rangle_{RR'}\,.
\end{equation}

When reasoning about decoding, it is convenient to refer to the given state of the world:
\begin{equation}
|\widetilde{\Psi}\rangle=\: \figbox{1.0}{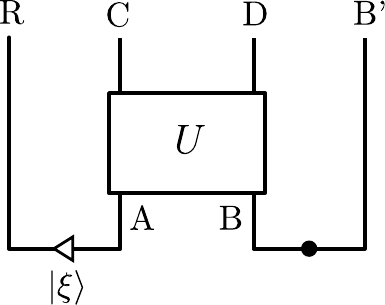}
\:,\qquad\qquad
\tilde{\rho}=|\widetilde{\Psi}\rangle\langle\widetilde{\Psi}|.\label{eq-world-state}
\end{equation}
We omit the tildes if $d_R=d_A$. Information-theoretically, the decoding is possible if the black hole has lost any memory of Alice's diary and become uncorrelated with $R$, i.e.\ if $\tilde{\rho}_{RC} \approx\tilde{\rho}_{R}\otimes\tilde{\rho}_{C}$. In particular, it is sufficient for $\tilde{\rho}_{RC}$ to be close to the maximally mixed state. This condition can be quantified using the parameter
\begin{equation}\label{delta}
\delta=d_{R}d_{C}\Tr\tilde{\rho}_{RC}^2-1\ge 0.
\end{equation}
If $\delta$ is small, then the decoding can be achieved with high fidelity, and our algorithms do exactly that. The number $\delta$ can be found from the diagrammatic expressions
\begin{equation}
\tilde{\rho}_{RC}=\: \frac{1}{d_B}\figbox{1.0}{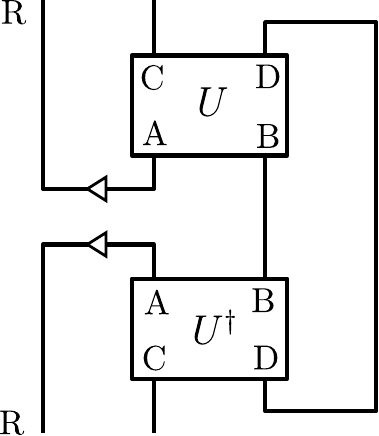}\:\,,
\qquad\quad
\Tr\tilde{\rho}_{RC}^2=\: \frac{1}{d_B^2}\figbox{1.0}{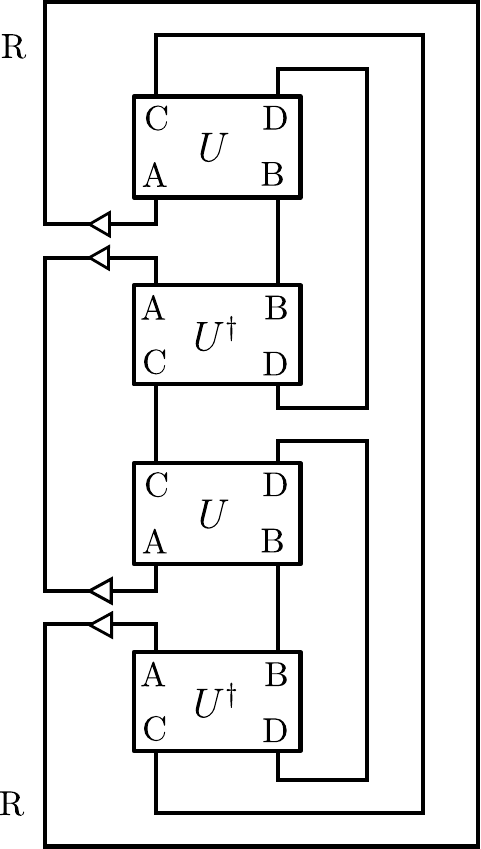}\:\,.
\end{equation}
After some rearrangement, we get this answer: 
\begin{equation}\label{Delta}
\delta =d_Ad_R\Delta-1,\qquad 
\text{where}\quad \Delta=\:\figbox{1.0}{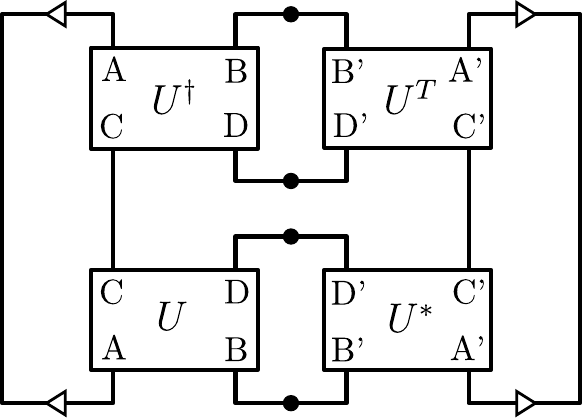}\:\,.
\end{equation}
If $d_R=d_A$, then $\Delta$ is also related to the R\'{e}nyi-$2$ mutual information between subsystems $R$ and $DB'$ for the density matrix $\rho$:
\begin{equation}\label{Delta_from_I2}
\Delta=2^{-I^{(2)}(R,DB')},\qquad\quad
I^{(2)}(R,DB')= S^{(2)}_{R} + S^{(2)}_{DB'} - S^{(2)}_{RDB'}\,.
\end{equation}
Here, $S^{(2)}_{DB'} =-\log_{2}\Tr(\rho_{DB'}^2) =-\log_{2}\Tr(\rho_{RC}^2)$ is the R\'{e}nyi-$2$ entropy of $DB'$ (equal to that of the complementary system $RC$), and it is clear that $S^{(2)}_{R}=\log_2d_A$ and $S^{(2)}_{RDB'}=S^{(2)}_{C}=\log_2d_C$.

\begin{figure}
\centering\includegraphics{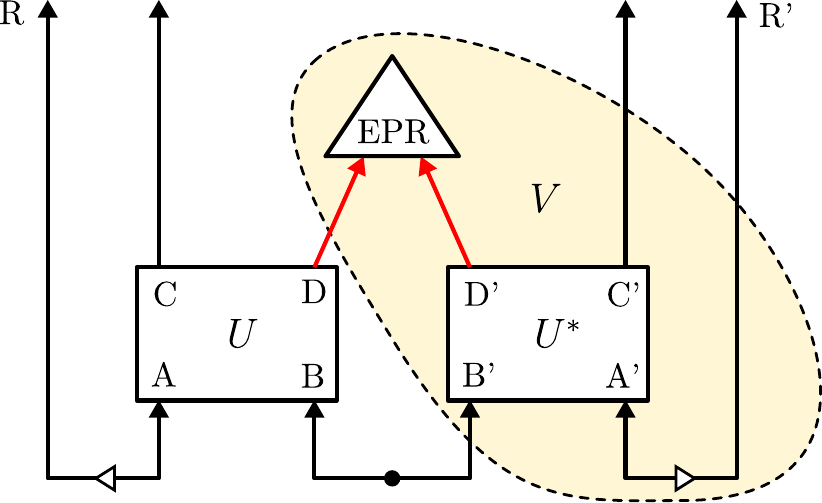}
\caption{Probabilistic decoding by postselecting on an EPR state. The actual decoder, denoted by $V$, corresponds to the shaded area.}
\label{Figure-probabilistic}
\end{figure}

We will construct two decoding algorithms that recover the entangled state $|\xi\rangle$ of the reference system and Alice's diary with fidelity $1-O(\delta)$. The bound $\delta\le d_Ad_R/d_D^2$ holds for a large class of operators $U$, hence, the algorithms work if $d_D\gg\sqrt{d_Ad_R}$. The first procedure successfully performs the task with probability $\approx\frac{1}{d_{A}d_{R}}$ or signals a failure. Roughly speaking, Bob tries to guess the content of Alice's diary. Classically, this is how one solves the equation $f(a)=r$ for an arbitrary function $f$: one guesses some candidate solution $a'$, calculates $r'=f(a')$, and compares it with $r$. In the quantum case, Bob prepares a copy of the entangled state $|\xi\rangle$ in separate subsystems $A'$, $R'$ and applies the operator $U^{*}$ to $A'B'$. Then he projects the captured Hawking radiation $D$ and its counterpart $D'$ onto the standard EPR state, see Figure~\ref{Figure-probabilistic}. If successful, the projection has the effect of ``teleporting'' Alice's part of $|\xi\rangle$ to subsystem $R'$. The second procedure is deterministic in the sense that it never aborts. It replaces the postselection with Grover's search, which involves applying $U^{*}$ and $U^{T}$ about $\sqrt{d_{A}d_{R}}$ times. Thus, this deterministic decoder has complexity $\mathcal{O}\bigl(\sqrt{d_{A}d_{R}}\,\mathcal{C}\bigr)$, where $\mathcal{C}$ is the complexity of implementing $U$. 

Now, let us discuss some physical assumptions that went into the definition of the problem. As already mentioned, we approximate the black hole thermal state by the maximally mixed state of dimension $d=2^S$. More exactly, each eigenvalue $Z^{-1}e^{-E_j/T}$ of the thermal density matrix is replaced by either $d^{-1}=Z^{-1}e^{-E/T}$ or $0$. This change is not extensive and amounts to neglecting energy fluctuations, which are relatively small in the thermodynamic limit. Yet the trace norm distance between the exact and approximate states is not small, therefore the decoding may work in one case but not the other. We believe that our algorithms can be adapted to a more realistic setting, though this involves some technical issues. At least, the probabilistic procedure generalizes to thermal states under the following assumptions. Let $\rho_{AB}$ be the thermal density matrix of the black hole that has absorbed the full energy of Alice's diary and let $\rho_{CD}=U\rho_{AB}U^\dag$. Our calculations work if $\rho_{AB}$ and $\rho_{CD}$ factor as $\rho_A\otimes\rho_B$ and $\rho_C\otimes\rho_D$, respectively. Unfortunately, this condition is problematic for physical reasons. Indeed, if $\rho_{CD}$ is thermal (which is almost true because the black hole evaporates adiabatically), the condition $\rho_{CD}=\rho_C\otimes\rho_D$ is equivalent to $H=H_C+H_D$. In other words, subsystems $C$ and $D$ do not interact! Of course, the Hawking radiation quanta eventually decouple from the black hole, but it is difficult to draw a sharp line between the two subsystems such that the density matrix factors. This issue can hopefully be addressed by soft partitioning of the subsystem $D$. After all, it is sufficient to verify that $D$ and $D'$ have enough entanglement, but we can be less aggressive in testing the qubits that are close to the boundary with $C$. Developing such a technique is a separate problem, and we set it aside.

\section{Condition on $U$ and out-of-time-order correlators}\label{sec_OTOC}

To obtain a good upper bound on the fidelity parameter $\delta$ or the related number $\Delta$ (see Eq.~(\ref{Delta})), we will assume that the evolution operator $U=e^{-iHt}$ is ``perfectly scrambling''. Quantum information scrambling is related to out-of-time-order correlators (OTOCs)~\cite{Shenker:2013pqa, Roberts:2014isa, Kitaev:2014t1, Maldacena:2016aa}. They have the general form $\corr{W(t)Y(0)Z(t)X(0)}$, where $O(0)=O$,\, $O(t)=U^{\dag}OU$, and the quantum average\footnote{\label{foot_OTOC}One may consider more general averages: $\Tr\bigl( W(t)\rho^{\alpha_3} Y(0)\rho^{\alpha_2} Z(t)\rho^{\alpha_1} X(0)\rho^{\alpha_0}\bigr)$, where $\alpha_3+\alpha_2+\alpha_1+\alpha_0=1$. In many cases, this change of the definition can be compensated by conjugating $X,Y,Z,W$ by suitable powers of $\rho$, or equivalently, by changing $Z(t)$ to $Z(t-i\alpha_1/T)$, etc. The most convenient choice is $\alpha_3=\alpha_2=\alpha_1=\alpha_0=1/4$.} of an operator $O$ is defined as $\langle O\rangle=\Tr O\rho$. In our case, $\rho=d^{-1}I_{AB}$, the operators $X$ and $Y$ act on subsystem $A$, whereas $Z$ and $W$ act on subsystem $D$. The (almost) perfect scrambling is defined as follows:
\begin{equation}\label{late-OTOC}
\bcorr{W(t)\,Y(0)\,Z(t)\,X(0)}
\approx \corr{WZ}\corr{Y}\corr{X}
+\corr{W}\corr{Z}\corr{YX}
-\corr{W}\corr{Z}\corr{Y}\corr{X}.
\end{equation}
This property holds for the Haar-random $U$, where the above equation becomes exact if we average the left-hand side over $U$ and subtract $(d^2-1)^{-1}\ccorr{WZ}\ccorr{YX}$ from the right-hand side. The derivation is given in Appendix~\ref{sec_averaging}. We work in the $d\to\infty$ limit; therefore, the correction just mentioned is negligible. More generally, equation~(\ref{late-OTOC}) characterizes the late-time asymptotics of OTOCs. It is expected to be true for a large class of operators $U$, provided $X$, $Y$, $Z$, $W$ act on sufficiently small subsystems.

If $d_R=d_A$, one can express $\Delta=2^{-I^{(2)}(R,DB')}$ using a formula derived in Refs.~\cite{Hosur:2015ylk,Roberts:2017aa}:
\begin{equation}
2^{-I^{(2)}(R,DB')} =\langle\text{OTOC}\rangle_{\text{ave}}.
\end{equation}
To handle the general case, we will devise some new notation. The idea is to combine $X$, $Y$ in the definition of OTOC into one object and $Z$, $W$ into another object:
\begin{equation}
\begin{aligned}
\OTOC(Y^T\otimes X,\,W\otimes Z^T)
&=\:\frac{1}{d}\Tr\bigl(U^{\dag}WU\,Y\,U^{\dag}ZU\,X\bigr)\\[5pt]
&=\: \frac{1}{d}\:\figbox{1.0}{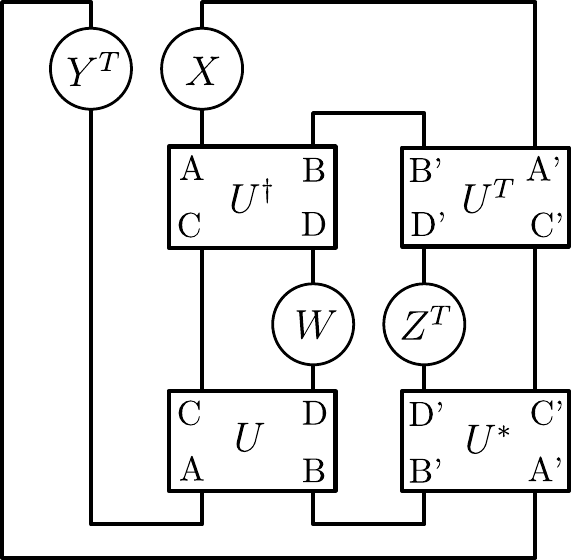}
\end{aligned}
\end{equation}
By linearity, this definition extends to arbitrary operators $L=\sum_{j}Y_j^T\otimes X_j$ and $M=\sum_{k}W_k\otimes Z_k^T$ acting on $A'A$ and $DD'$, respectively. In this notation,
\begin{equation}
\Delta=\OTOC(L,M),\qquad \text{where}\quad\:
L=d_A\,\figbox{1.0}{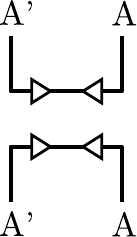}\:,\qquad
M=\figbox{1.0}{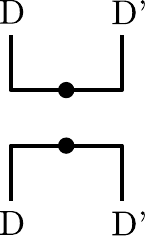}\:.
\end{equation}

Now, let us derive a special case of Eq.~(\ref{late-OTOC}) for $\OTOC(L,M)$. The left-hand side of the original equation is equal to $\OTOC(Y^T\otimes X,\,W\otimes Z^T)$. The right-hand side contains these linear functions of $Y^T\otimes X$:
\begin{equation}
\corr{Y}\corr{X}=\,\figbox{1.0}{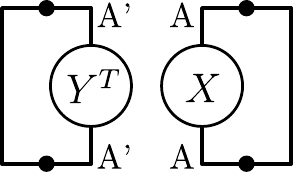}\,,\qquad\quad
\corr{YX}=\,\figbox{1.0}{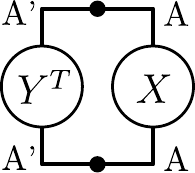}\,,
\end{equation}
as well as similar expressions with $W$ and $Z$. We can apply the same graphics rules to $L$ and $M$ and evaluate the resulting diagrams. Thus, if $U$ is almost perfectly scrambling, then
\begin{equation}
\OTOC(L,M)\approx \frac{1}{d_Ad_R}+\frac{1}{d_D^2}-\frac{1}{d_Ad_Rd_D^2}.
\end{equation}
Assuming that the error in this approximation is smaller than the last term, we conclude that
\begin{equation}
\Delta \le \frac{1}{d_Ad_R}+\frac{1}{d_D^2},\qquad
\delta=d_Ad_R\Delta-1\le \frac{d_Ad_R}{d_D^2}.
\end{equation}

\section{Probabilistic decoder}\label{sec:protocol}

The probabilistic decoding begins with the state $|\widetilde{\Psi}\rangle$ defined in Eq.~\eqref{eq-world-state}. As a very first step, Bob creates a copy of $|\xi\rangle$ on $A'R'$ and applies $U^{*}$. The result is:
\begin{align}
|\Psi_{\text{in}}\rangle = (I_{RC}\otimes I_{D}\otimes U^{*}\otimes I_{R'}) ( |\widetilde{\Psi}\rangle \otimes |\overleftrightarrow{\xi} \rangle )
= \: \figbox{1.0}{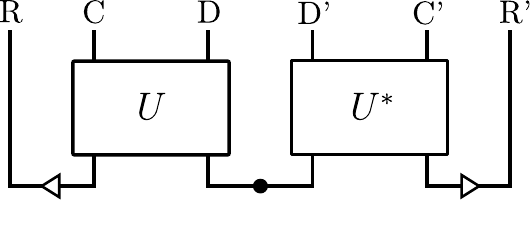}\:, \label{eq-in-state}
\end{align}
where $|\overleftrightarrow{\xi} \rangle$ is obtained by swapping $R$ and $A$ in $|\xi \rangle$. Then Bob applies a projector $P_{D}$ onto the EPR pair on $DD'$. It is defined as an operator acting only on $DD'C'R'$ because $R$ and $C$ are not accessible by Bob:
\begin{align}
P_{D}= \bigl(|\text{EPR}\rangle_{DD'}\langle\text{EPR}|_{DD'}\bigr) \otimes I_{C'R'} = \: \figbox{1.0}{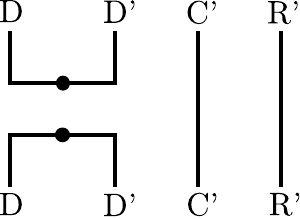}\:.
\end{align}
The projection succeeds with probability $\langle\Psi_{\text{in}}| I_{RC}\otimes P_{D} |\Psi_{\text{in}}\rangle =\Delta$, where $\Delta$ is given by Eq.~(\ref{Delta}). Note that $\Delta\ge\frac{1}{d_{A}d_{R}}$ because $\delta\ge 0$. The normalized output state is
\begin{align}
|\Psi_{\text{out}}\rangle
= \frac{1}{\sqrt{\Delta}} (I_{RC}\otimes P_{D}) |\Psi_{\text{in}}\rangle
= \: \frac{1}{\sqrt{\Delta}} \:\figbox{1.0}{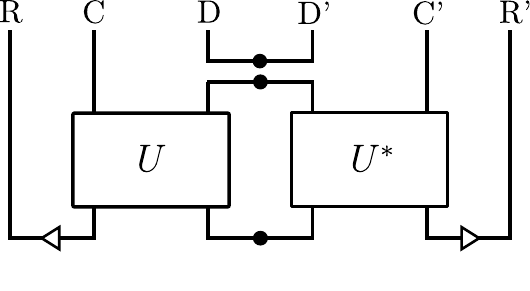}\:.
\end{align}
By definition, the decoding is exact if $|\Psi_{\text{out}}\rangle$ contains the EPR pair on $RR'$. In general, we do not expect it to be exact and should estimate the associated fidelity.

We will first argue abstractly that if the Hayden-Preskill decoding is information-theoretically possible, then our probabilistic decoder also works. This argument is not meant as a rigorous proof, but it provides some insight and reveals the connection between our scheme and quantum teleportation. A more concrete analysis will follow. 

Suppose that a perfect decoder $V$ exists that works as shown in Figure \ref{fig_Hayden_Preskill_decoding}b. Then for any operator $X_R$ there is some operator $X_{DB'}$ (namely, $X_{DB'}=V^{\dag}(I_{E}\otimes X_{R}^T)V$) such that
\begin{align}
\figbox{1.0}{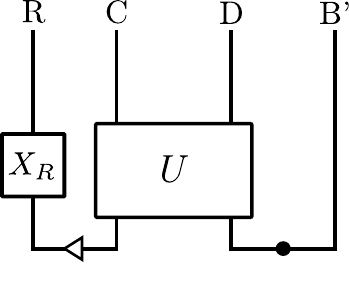} = \: \figbox{1.0}{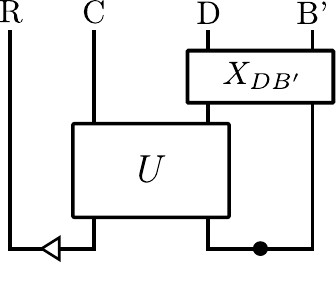}.
\end{align}
Furthermore, if $X_R$ is unitary, then $X_{DB'}$ is also unitary. A similar argument holds for $U^{*}$ by taking complex conjugates. Hence,
\begin{align}\label{fig_teleportation}
\figbox{1.0}{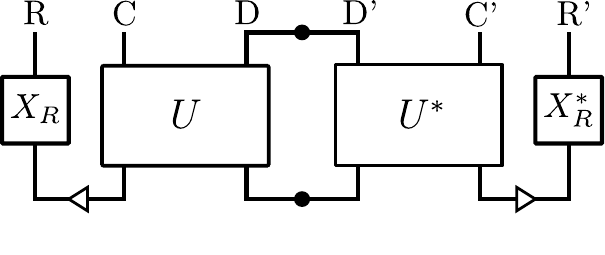} \: = \: \figbox{1.0}{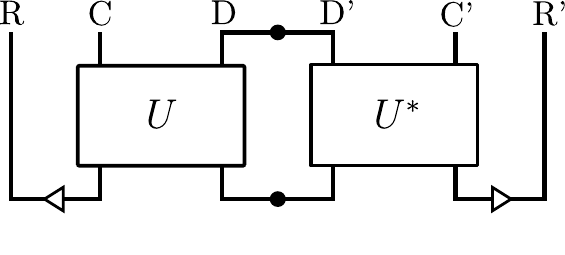}. 
\end{align}
Here, we have glued the diagrams with $U$ and $U^*$ together, removed one dot from the bottom and added it on the top. The right-hand side is, essentially, the state $|\Psi_{\text{out}}\rangle$ up to trivial changes. The whole equation is equivalent to the condition that $(X_{R}\otimes I_{CDD'C'}\otimes X^*_{R'})|\Psi_{\text{out}}\rangle =|\Psi_{\text{out}}\rangle$ for all unitaries $X_R$. It follows that $|\Psi_{\text{out}}\rangle$ contains an EPR pair on $RR'$, and thus, fulfills the decoding requirement.

When $R$, $A$, $D$, $D'$, $A'$, $R'$ are single qubits, the probabilistic decoding has the effect of ``teleporting'' Alice's quantum state to subsystem $R'$ by postselecting on a particular Bell state. While ordinary quantum teleportation succeeds even if the Bell measurement outcome is different, in the probabilistic decoder, projections onto wrong Bell basis states imply failure of decoding in general as $R'$ may remain entangled with $CC'$.

The fidelity of the probabilistic decoder is 
\begin{equation}
F = \langle \Psi_{\text{out}}|P_{R}|\Psi_{\text{out}}\rangle
=\frac{\langle \Psi_{\text{in}} |  P_{R} (I_{RC}\otimes P_{D}) | \Psi_{\text{in}}\rangle }{\Delta}, \qquad \text{where}\quad\:
P_{R} = \:\figbox{1.0}{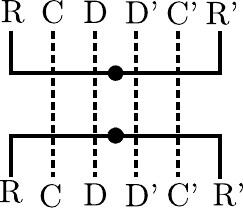}.
\end{equation}
The numerator in the expression for $F$ can be lower bounded by $\bigl|\langle\text{EPR}|_{RCD}| \Psi_{\text{in}}\rangle\bigr|^2$, where
\begin{equation}\label{EPR_Psi_in}
\begin{split}
\langle \text{EPR} |_{RCD} | \Psi_{\text{in}}\rangle &= \: \figbox{1.0}{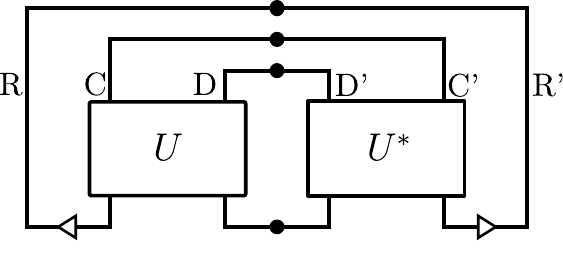}\\
&= \: \figbox{1.0}{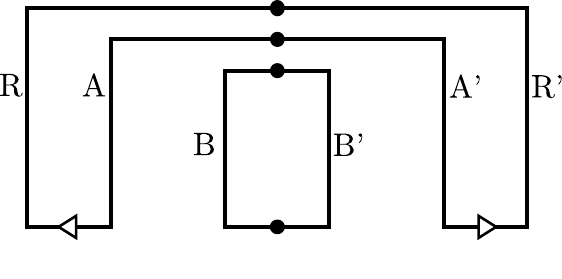} \: = \frac{1}{\sqrt{d_{R}d_{A}}}.
\end{split}
\end{equation}
Therefore, 
\begin{equation}
F \geq \frac{1}{d_{R}d_{A}\Delta} = \frac{1}{1+\delta}.
\end{equation}
We conclude that if $U$ is almost perfectly scrambling and $d_{D}\gg \sqrt{d_{A}d_{R}}$, then $F \approx 1 $.\medskip

The above calculation generalizes to factorizable inputs and outputs by replacing each dot with the square root of the corresponding density matrix. (Note that $\rho_R$ is still maximally mixed.) Using this definition and the condition $\rho_C\otimes\rho_D=U(\rho_A\otimes\rho_B)U^{\dag}$, the dots on the $C$ and $D$ lines in the first diagram in~(\ref{EPR_Psi_in}) can be moved through $U$. Since the dot on the $R$ line corresponds to the factor $d_R^{-1/2}$ and each triangle to the operator $d_R^{-1/2}\Xi$, we arrive at the following bound:
\begin{equation}
F\ge \frac{d_R^{-3}
\bigl(\Tr(\Xi^{\dag}\sqrt{\rho_A}\,\Xi)\bigr)^2}{\Delta},
\end{equation}
The expressions for OTOCs in Section~\ref{sec_OTOC} are generalized by interspersing $WYZX$ with $\rho^{1/4}$ as described in footnote~\ref{foot_OTOC}. Expectation values like $\corr{X}$ are defined with respect to the density matrix, whereas $\corr{XY}$ involves two copies of $\sqrt{\rho}$. If $U$ is almost perfectly scrambling, then
\begin{equation}
\Delta\ge \frac{1}{\widetilde{d_A}d_R}+\frac{1}{\widetilde{d_D}^2},\qquad
\text{where}\quad
\widetilde{d_A}
=\bigl(d_R^{-1}\Tr(\Xi^{\dag}\sqrt{\rho_A}\,\Xi)^{2}\bigl)^{-1},\quad\:
\widetilde{d_D}
=\bigl(\Tr\rho_D^3\bigr)^{-1/2}.
\end{equation}
In this case, the fidelity bound becomes
\begin{equation}
F\ge \frac{\bigl(d_R^{-1}\Tr(\Xi^{\dag}\sqrt{\rho_A}\,\Xi)\bigr)^2}
{d_R^{-1}\Tr(\Xi^{\dag}\sqrt{\rho_A}\,\Xi)^{2}}\,
\frac{1}{1+\widetilde{d_A}d_R/\widetilde{d_D}^2}.
\end{equation}
Qualitatively, the first factor is close to $1$ if the image of the embedding $\Xi$ lies in the ``typical subspace'' of $\rho_A$.

\section{Deterministic decoder}\label{sec:Grover}

Unfortunately, the success probability of the aforementioned decoding algorithm scales as $\frac{1}{d_{A}d_{R}}$. Now we present a modified decoder that is deterministic (albeit not exact). It incorporates a procedure similar to Grover's search algorithm~\cite{Grover:1996}. 

The initial step is the same as in the probabilistic decoding so that the subsequent procedure is applied to the state $|\Psi_{\text{in}}\rangle$ defined by Eq.~\eqref{eq-in-state}. Let $P_{A}$ be the projector onto $|\overleftrightarrow{\xi}\rangle$ on $A'R'$ and define another projector $\widetilde{P}_{A}=(I_{D}\otimes U^{*}\otimes I_{R'})P_{A}(I_{D}\otimes U^{T}\otimes I_{R'})$ acting on $DD'C'R'$:
\begin{align}
P_{A} = \figbox{1.0}{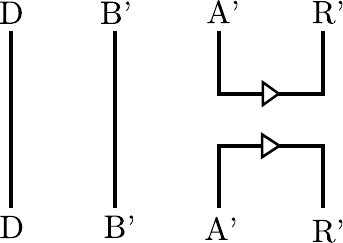}, \qquad\quad \widetilde{P_{A}} = \figbox{1.0}{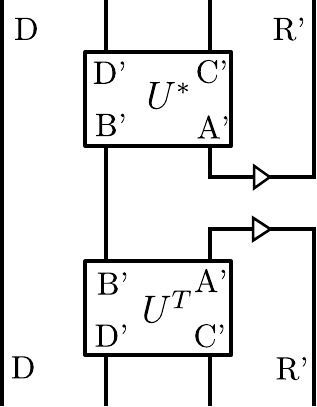}\,\:.
\end{align}
Consider the following unitary operators:
\begin{align}
W_{D} = 1 - 2P_{D}, \qquad\quad
\widetilde{W}_{A} = 2\widetilde{P}_{A} - 1.
\end{align}
Bob's decoding strategy is to apply $W=\widetilde{W}_{A}W_{D}$ multiple ($\approx \frac{\pi \sqrt{d_{A}d_{R}} }{4}$) times to obtain a good approximation of $|\Psi_{\text{out}}\rangle$. 

To analyze the algorithm, it will be convenient to define the following operator acting on subsystem $DD'C'R'$:
\begin{align}
\Pi \equiv \widetilde{P}_{A}P_{D}\widetilde{P}_{A} = \: \figbox{1.0}{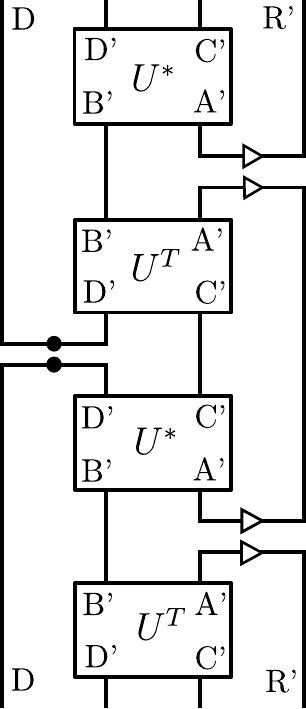}
\end{align}
The rank $r$ of this operator is at most the rank of $P_{D}$, which is $d_{R}d_{C}$. The following identity can be derived graphically by rotating the $U^T$ and $U^{*}$ boxes in the middle in opposite directions and moving them to the left:
\begin{align}
\Tr_{RC}(|\Psi_{\text{in}}\rangle \langle   \Psi_{\text{in}}|) = \: \figbox{1.0}{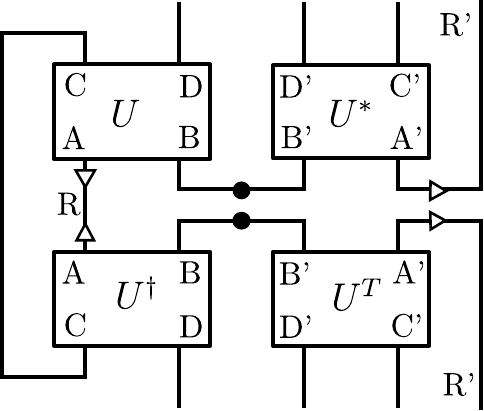}\: =\: \frac{d_{A}}{d_{C}} \Pi. \label{eq:trace}
\end{align}
Equation~\eqref{eq:trace} helps to decompose the initial vector $|\Psi_{\text{in}}\rangle$ into eigenvectors of $I_{RC}\otimes \Pi$. To this end, let us consider the eigendecomposition of $\Pi$:
\begin{align}
\Pi = \sum_{j=1}^{r}\alpha_{j} |\psi_{j}\rangle\langle \psi_{j}|, \qquad \alpha_{j}>0.
\end{align}
The vectors $|\psi_{j}\rangle$ are also eigenvectors of $\Tr_{RC}(|\Psi_{\text{in}}\rangle \langle\Psi_{\text{in}}|)$. Together with some orthonormal vectors $|\eta_{j}\rangle$ on the complementary subsystem $RC$, they make a Schmidt decomposition of $|\Psi_{\text{in}}\rangle$:
\begin{align}\label{Psi_j}
|\Psi_{\text{in}}\rangle
= \sum_{j=1}^{r} \sqrt{\frac{d_{A}}{d_{C}}}\,
\sqrt{\alpha_{j}}\, |\Psi_{j}\rangle,\qquad\quad
|\Psi_{j}\rangle = |\eta_{j}\rangle \otimes |\psi_{j}\rangle.
\end{align}
Since $\langle \Psi_{\text{in}} | \Psi_{\text{in}} \rangle$=1 and $\langle \Psi_{\text{in}} | \Pi | \Psi_{\text{in}} \rangle = \Delta$, we have 
\begin{align}
\sum_{j=1}^{r}\alpha_{j}=\frac{d_{C}}{d_{A}}, \qquad\quad \sum_{j=1}^{r}\alpha_{j}^2= \frac{d_{C}}{d_{A}}\,\Delta, \qquad
\text{where}\quad r\le d_Rd_C. \label{eq:average}
\end{align}
In the ideal case of $\Delta=1/(d_{A}d_{R})$, these conditions imply that $\alpha_{j}=1/(d_{A}d_{R})$ for all $j$ and that $r=d_Rd_C$.

Now we examine how the iterated application of $I_{RC}\otimes W$ acts on $|\Psi_{\text{in}}\rangle$. Let us define the following unit vectors:
\begin{equation}
|\Phi_{j}\rangle
= \frac{I_{RC}\otimes \widetilde{P}_{D}}{\sqrt{\alpha_{j}}}
|\Psi_{j}\rangle. 
\end{equation}
It follows from equations below that the two-dimensional subspace $\mathcal{L}_j=\text{linear span}\{|\Psi_{j}\rangle, |\Phi_{j}\rangle\}$ is invariant under both $I_{RC}\otimes\widetilde{P}_{A}$ and $I_{RC}\otimes P_{D}$. Furthermore, $\mathcal{L}_j\perp\mathcal{L}_k$ if $j\not=k$. Thus, for each individual $j$, our procedure works as the standard Grover algorithm. We will reproduce its analysis for reader's convenience and then apply it to a superposition of $j$'s.
\begin{figure}
\centering\(\displaystyle
\begin{array}{@{}c@{\hspace{2cm}}c@{}}
\figbox{1.0}{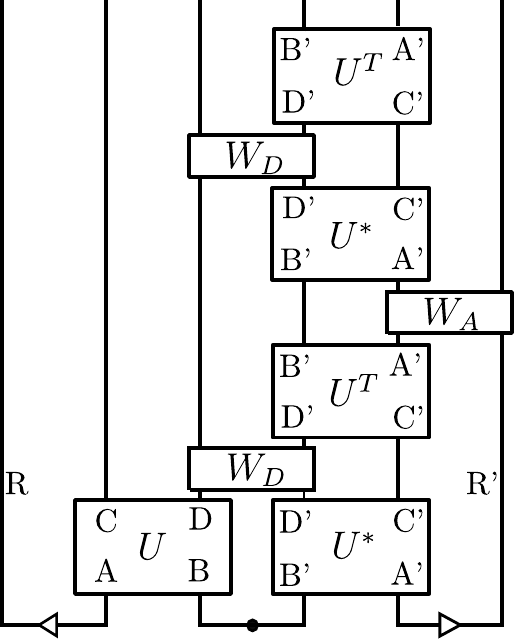} &
\figbox{1.0}{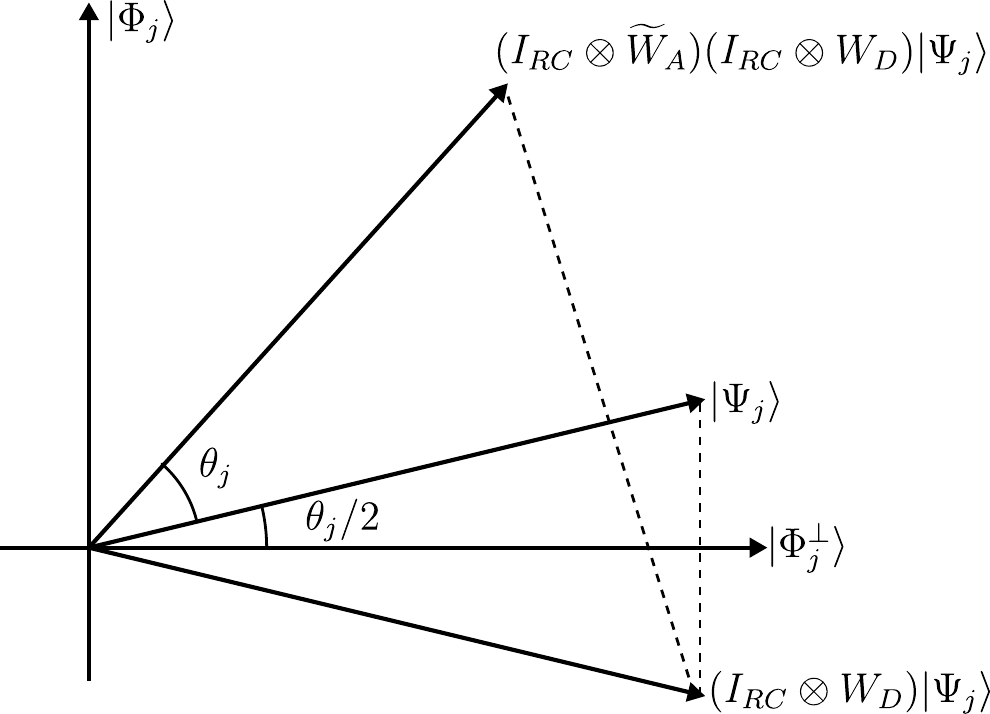}
\vspace{5pt}\\
\text{a)} & \text{b)}
\end{array}
\)
\caption{The deterministic decoder (a) and the Grover rotation (b).}
\label{Figure-Grover}
\end{figure}

Inside $\mathcal{L}_{j}$, the algorithm induces Grover rotations by some angle that depends on $\alpha_{j}$. To see this, we will use the following relations:
\begin{align}
\label{PsiPhi1}
I_{RC}\otimes P_{D}|\Psi_{j}\rangle &=\sqrt{\alpha_{j}}|\Phi_{j}\rangle, \qquad
&I_{RC}\otimes P_{D}|\Phi_{j}\rangle &= |\Phi_{j}\rangle, \\[3pt]
\label{PsiPhi2}
I_{RC}\otimes \widetilde{P}_{A}|\Psi_{j}\rangle &= |\Psi_{j}\rangle, \qquad
&I_{RC}\otimes \widetilde{P}_{A} |\Phi_{j}\rangle
&= \sqrt{\alpha_{j}}|\Psi_{j}\rangle.
\end{align}
The first pair of equations follows from the definition of $|\Phi_{j}\rangle$ and the second from the fact that $|\Psi_{j}\rangle$ is an eigenvectors of $I_{RC}\otimes \Pi$ with eigenvalue $\alpha_{j}$. The vector $|\Psi_{j}\rangle$ can be represented as $\sin(\theta_j/2)\,|\Phi_{j}\rangle +\cos(\theta_j/2)\,|\Phi_{j}^{\perp}\rangle$, where $|\Phi_{j}^{\perp}\rangle\in\mathcal{L}_{j}$ is a unit vector orthogonal to $|\Phi_{j}\rangle$, see Figure~\ref{Figure-Grover}. The value of $\theta_j$ can be obtained from Eq.~(\ref{PsiPhi2}):
\begin{align}
\sin(\theta_{j}/2) = \langle\Phi_j|\Psi_j\rangle = \sqrt{\alpha_{j}}.
\end{align}
Notice that $I_{RC}\otimes W_{D}$ is a reflection across $|\Phi_{j}^{\perp}\rangle$ and $I_{RC}\otimes \widetilde{W}_{A}$ is a reflection across $|\Psi_j\rangle$. Therefore, the operator $I_{RC}\otimes W=(I_{RC}\otimes \widetilde{W}_{A})(I_{RC}\otimes W_{D})$ restricted to $\mathcal{L}_{j}$ is a rotation by angle $\theta_{j}$; its action on the vector $|\Psi_j\rangle$ is shown in the figure. Now recall that the initial vector $|\Psi_{\text{in}}\rangle$ is a superposition of $|\Psi_j\rangle$ with different $j$, which is given by Eq.~(\ref{Psi_j}). After $m$ iterations, the quantum state becomes
\begin{align}
|\Psi(m)\rangle
= \sum_{j=1}^{r} \sqrt{\frac{d_{A}}{d_{C}}}\, \sqrt{\alpha_{j}}\,
\Bigl(\sin\bigl(\bigl(m+\tfrac{1}{2}\bigr)\theta_{j}\bigr)\,|\Phi_{j}\rangle
+\cos\bigl(\bigl(m+\tfrac{1}{2}\bigr)\theta_{j}\bigr)\,|\Phi_{j}^{\perp}\rangle
\Bigr).  \label{eq:rotated}
\end{align}

In the ideal case, we would have $\theta_j=\theta_*=2\arcsin\bigl((d_Ad_R)^{-1/2}\bigr)$ for all $j$. Setting $m$ to
\begin{equation}
m_*=\frac{\pi}{2\theta_*}-\frac{1}{2} \approx\frac{\pi}{4}\,\sqrt{d_Ad_R}
\end{equation}
would give the state $|\Psi(m_*)\rangle= \sum_{j=1}^{r}\sqrt{\frac{d_{A}}{d_{C}}}\,
\sqrt{\alpha_{j}}\,|\Phi_{j}\rangle$, which can be shown to satisfy the decoding requirement. Let us bound the fidelity of the reconstructed state $|\Psi(m_*)\rangle$ under more general circumstances. To simplify the analysis, we assume that $d_{A}d_{R}\gg 1$ so that the error due to the rounding of $m_*$ to an integer may be neglected.\footnote{Instead of rounding, one can alter the last Grover step so as to rotate the vector by a suitable angle less than $\theta_j$. For example, one can use an operator of the form $\bigl(e^{i\beta}(1-\widetilde{P}_A)+\widetilde{P}_A\bigr) \bigl((1-P_D)+e^{i\gamma}P_D\bigr)$. Thus modified algorithm should be accurate even if $d_{A}d_{R}\sim 1$.} We will also approximate $\theta_j =2\arcsin\sqrt{\alpha_j}$ by $2\sqrt{\alpha_j}$; the resulting error is also suppressed when $d_{A}d_{R}\gg 1$.

Let us upper bound the distance between $|\Psi(m_*)\rangle$ and a state that is known to have $1-O(\delta)$ decoding fidelity, namely
\begin{align}
|\Psi_{\text{out}}\rangle
= \frac{1}{\sqrt{\Delta}} (I_{RC}\otimes P_{D}) |\Psi_{\text{in}}\rangle
= \frac{1}{\sqrt{\Delta}} \sum_{j=1}^{r}
\sqrt{\frac{d_{A}}{d_{C}}}\,\alpha_{j}|\Phi_{j}\rangle.
\end{align}
Consider the following auxiliary vector:
\begin{equation}
|\Psi_{\text{out}}'\rangle
= \sum_{j=1}^{r}\sqrt{\frac{d_{A}}{d_{C}}}\,
\sqrt{\alpha_{j}}\,|\Phi_{j}\rangle. 
\end{equation}
It is sufficiently close to $|\Psi_{\text{out}}\rangle$. Indeed,
\begin{equation}
\begin{alignedat}{2}
\bigl\| |\Psi_{\text{out}}\rangle-|\Psi_{\text{out}}'\rangle \bigr\|^2
&= \frac{d_{A}}{d_{C}} \sum_{j=1}^{r}  \alpha_{j} \left(\frac{\sqrt{\alpha_{j}}}{\sqrt{\Delta}} -1 \right)^2
&& \\
&\leq
\frac{d_{A}}{d_{C}} \sum_{j=1}^{r}  \alpha_{j} \left(\frac{\sqrt{\alpha_{j}}}{\sqrt{\Delta}} - \frac{\sqrt{\Delta}}{\sqrt{\alpha_{j}}} \right)^2
\qquad && \text{since } |x-1|\le|x-x^{-1}| \text{ for } x>0 \\
&= \frac{d_{A}}{d_{C}} \sum_{j=1}^{r}
\left( \frac{\alpha_{j}^2}{\Delta} - 2 \alpha_{j} + \Delta\right)
&& \\
&\le d_{A}d_{R}\Delta - 1 = \delta
\qquad && \text{using Eq.~(\ref{eq:average})}.
\end{alignedat}
\end{equation}
Let $x_{j}=\sqrt{\alpha_{j}d_{A}d_{R}}$ so that $\bigl(m_*+\frac{1}{2}\bigr)\theta_j\approx\frac{\pi}{2}x_j$ and thus,
\begin{align}
|\Psi(m_*) \rangle \approx \sum_{j=1}^{r} 
\sqrt{\frac{d_{A}}{d_{C}}} \sqrt{\alpha_{j}}
\Bigl(\sin \bigl(\tfrac{\pi}{2}x_{j}\bigr) |\Phi_{j}\rangle + \cos\bigl(\tfrac{\pi}{2}x_{j}\bigr)|\Phi_{j}^{\perp}\rangle \Bigr).
\end{align}
The distance between $|\Psi(m_*)\rangle$ and $|\Psi_{\text{out}}'\rangle$ can be bounded as follows:
\begin{equation}
\begin{split}
\bigl\| |\Psi(m_*)\rangle-|\Psi_{\text{out}}'\rangle \bigr\|^2
&\approx \frac{d_{A}}{d_{C}} \sum_{j=1}^{r} 2\alpha_{j}
\biggl( 1 - \cos\Bigl(\frac{\pi}{2} (x_{j}-1) \Bigr) \biggr)\\
&\leq \frac{d_{A}}{d_{C}}\Big( \frac{\pi}{2} \Big)^2  \sum_{j=1}^{r}  \alpha_{j} (x_{j}-1)^2  \\
&\leq \frac{d_{A}}{d_{C}}\Big( \frac{\pi}{2} \Big)^2  \sum_{j=1}^{r}  \alpha_{j} \Big(x_{j}-\frac{1}{x_{j}}\Big)^2 
\le \Big( \frac{\pi}{2} \Big)^2 \delta,
\end{split}
\end{equation}
where Eq.~(\ref{eq:average}) was used at the last step. Therefore,
\begin{equation}
\bigl\| |\Psi(m_*)\rangle-|\Psi_{\text{out}}\rangle \bigr\|
\le \left(1+\frac{\pi}{2}\right)\sqrt{\delta}\,,
\end{equation}
which shows that $|\Psi(m_*)\rangle$ satisfies the decoding condition with $1-O(\delta)$ fidelity.\medskip

The above argument generalizes to factorizable inputs and outputs if we also assume that the image of $\Xi$ is an invariant subspace of $\rho_A$. The parameter $\delta$ in the fidelity bound should be replaced with
\begin{equation}
\widehat{\delta}=\widehat{d_{A}}d_R\Delta-1,\qquad \text{where}\quad
\widehat{d_{A}} = d_{R}^{-1} \Tr\bigl(\Xi^{\dagger}\rho_{A}^{-1}\Xi\bigr).
\end{equation}
We leave this generalization as an exercise for reader.

\section{Discussion}\label{sec:discussion}

Our decoding procedures bear some similarity with the Gao-Jafferis-Wall traversable wormhole, but differ in some aspects. First, the Gao-Jafferis-Wall system achieves deterministic decoding in one go, without Grover-like iterations. However, the signal can only cross the wormhole if it is sent within a certain time window; therefore, the simplicity of decoding relies on some properties of the operator $U$ at early times. It would be interesting to find exactly what these properties are. Second, the wormhole is just a variant of an eternal black hole. As such, it is characterized by a thermal state or its thermofield double rather than the maximally mixed/entangle state. Our algorithms can be adapted to a setting where $\rho_{AB}$ and $\rho_{CD}$ need not be maximally mixed but factor as $\rho_{A}\otimes \rho_{B}$ and $\rho{_C}\otimes\rho_{D}$. This assumption is unrealistic though, and finding a good generalization to thermal states is still an open problem.

The deterministic decoding algorithm with Grover iterations can be expressed as higher-order OTOCs of the form $\langle X_{1}(0)Y_{1}(t)X_{2}(0)Y_{2}(t) \cdots \rangle$. For instance, the expectation value of $\Pi^m = \tilde{P}_{A}  (P_{D} \tilde{P}_{A} )^m $ with respect to the initial state $|\Psi_{\text{in}}\rangle$ can be expressed in terms of $4m$-point OTOCs. Interestingly, this expectation value, $\langle \Psi_{\text{in}} | \Pi^m|\Psi_{\text{in}}\rangle$,  can be computed from the R\'{e}nyi-$2m$ mutual information $I^{(2m)}(R,DB')$. The type of higher-order OTOCs associated with our deterministic decoding algorithm are similar to those previously considered by Shenker and Stanford~\cite{Shenker:2013yza} in the context of multiple shockwave geometries. It would be interesting to develop holographic and geometric interpretations of the deterministic decoding protocol. 

Another question concerns the optimality of our scheme. The Grover algorithm is known to be optimal for the black box search problem in the sense of query complexity~\cite{BBBV}. It would be interesting to see whether our deterministic decoder uses the operator $U$, considered as a black box, as few times as possible.

\section*{Acknowledgments}

We thank John Preskill, Daniel Roberts and Douglas Stanford for useful discussions. We gratefully acknowledge the support by the Simons Foundation through the ``It from Qubit'' program. Research at Perimeter Institute is supported by the Government of Canada through Industry Canada and by the Province of Ontario through the Ministry of Research and Innovation. A.K.\ is supported by the Simons Foundation under grant~376205 and by the Institute of Quantum Information and Matter, a NSF Frontier center funded in part by the Gordon and Betty Moore Foundation.

\appendix

\section{Averaging over the unitary group}\label{sec_averaging}

Let $X$ be an operator acting on $n$ qudits (i.e.\ systems with a $d$-dimensional Hilbert space). Its average over the unitary group $\UU(d)$ is defined as follows:
\begin{equation}
T_n(X)= \int U^{\otimes n}X(U^{\otimes n})^\dag\, dU,
\end{equation}
where the integral is taken over the Haar measure. For $n=2$, it is given by
\begin{equation}\label{T2}
T_2(X)=\frac{1}{d^2-1}\left(
I\,\Tr X+S\,\Tr SX-\frac{1}{d}\,S\,\Tr X-\frac{1}{d}\,I\,\Tr SX
\right)
\end{equation}
where $S$ is the qudit swap operator. Let us also write this using indices:
\begin{equation}
(T_2(X))^{j_1j_2}_{m_1m_2}
=\sum_{k_1,k_2,l_1,l_2}
(T_2)^{j_1}_{m_1} \mkern-1mu{}^{j_2}_{m_2}\mkern-1mu {}^{l_1}_{k_1}\mkern-1mu {}^{l_2}_{k_2}X^{k_1k_2}_{l_1l_2}.
\end{equation}
It follows from Eq.~(\ref{T2}) that
\begin{equation}
(T_2)^{j_1}_{m_1} \mkern-1mu{}^{j_2}_{m_2}\mkern-1mu {}^{l_1}_{k_1}\mkern-1mu {}^{l_2}_{k_2}
=\frac{1}{d^2-1}\left(
\delta^{j_1}_{m_1}\delta^{j_2}_{m_2}\delta^{l_1}_{k_1}\delta^{l_2}_{k_2}
+\delta^{j_1}_{m_2}\delta^{j_2}_{m_1}\delta^{l_1}_{k_2}\delta^{l_2}_{k_1}
-\frac{1}{d}\,
\delta^{j_1}_{m_2}\delta^{j_2}_{m_1}\delta^{l_1}_{k_1}\delta^{l_2}_{k_2}
-\frac{1}{d}\,
\delta^{j_1}_{m_1}\delta^{j_2}_{m_2}\delta^{l_1}_{k_2}\delta^{l_2}_{k_1}
\right).
\end{equation}
In graphic notation, this equation reads:
\begin{equation}
\figbox{1.0}{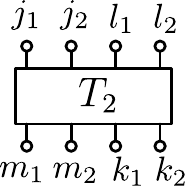}\: = \:
\frac{1}{d^2-1}\Biggl(
\figbox{1.0}{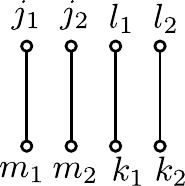}\,+\,\figbox{1.0}{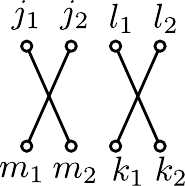}\,
-\frac{1}{d}\,\figbox{1.0}{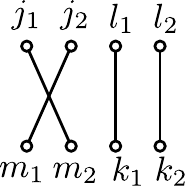}\,-\frac{1}{d}\,\figbox{1.0}{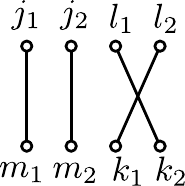}
\Biggr).
\end{equation}

Perhaps the most useful form of Eq.~(\ref{T2}) is an expression for an out-of-time-order correlator. For an arbitrary operator acting on just one subsystem, $X\in \LL(\mathbb{C}^d)$, let $\widetilde{X}=U^{\dag}XU$. The quantum average of $X$ over the maximally mixed state $\rho=\frac{I}{d}$ is $\corr{X}=\Tr X\rho=\frac{1}{d}\Tr X$. In addition, we average over the random unitary $U$, which is indicated by an overbar. 
Let $X,Y,Z,W\in \LL(\mathbb{C}^d)$, and let us calculate the following quantity:
\begin{align*}
d\:\overline{\bcorr{\widetilde{W}Y\widetilde{Z}X}}
\hspace{-20pt}&\hspace{20pt}
=\int\Tr\bigl((U^{\dag}WU)Y(U^{\dag}ZU)X\bigr)\,dU
=\int\Tr\bigl(W(UYU^{\dag})Z(UXU^{\dag})\bigr)\,dU\\[2pt]
&=\:\figbox{1.0}{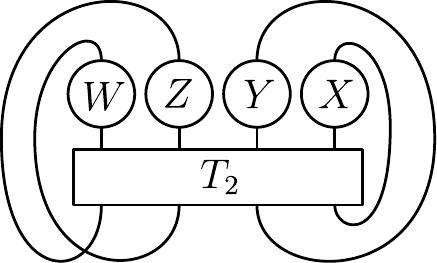}\:=\frac{1}{d^2-1}
\left(\begin{aligned}
&(\Tr WZ)(\Tr Y)(\Tr X)\\[1pt]
+&(\Tr W)(\Tr Z)(\Tr YX)\\
-&\frac{1}{d}(\Tr W)(\Tr Z)(\Tr Y)(\Tr X)\\
-&\frac{1}{d}(\Tr WZ)(\Tr YX)
\end{aligned}\right).
\end{align*}
After some arithmetic manipulation, we arrive at the following result:
\begin{equation}\label{OTOcorr}
\begin{aligned}
\overline{\bcorr{\widetilde{W}Y\widetilde{Z}X}}
&:=\int\bcorr{(U^{\dag}WU)Y(U^{\dag}ZU)X}\,dU\\[2pt]
&=\corr{WZ}\corr{Y}\corr{X}
+\corr{W}\corr{Z}\corr{YX}
-\corr{W}\corr{Z}\corr{Y}\corr{X}
-\frac{1}{d^2-1}\ccorr{WZ}\ccorr{YX},
\end{aligned}
\end{equation}
where
\begin{equation}
\ccorr{AB}:=\corr{AB}-\corr{A}\corr{B}
=\bcorr{(A-I\corr{A})(B-I\corr{B})}.
\end{equation}
When $d$ is large, the last term in Eq.~(\ref{OTOcorr}) may be neglected. Furthermore, if the quantum averages of $X$, $Y$, $Z$, $W$ are equal to zero (which is the case if we replace them by $X-I\corr{X}$, etc.), then the whole correlator vanishes. This is the qualitative effect of a random unitary --- to erase nontrivial out-of-time-order correlators.

\mciteSetMidEndSepPunct{}{\ifmciteBstWouldAddEndPunct.\else\fi}{\relax}
\bibliographystyle{utphys}

\begin{mcitethebibliography}{10}

\bibitem{Page93}
D.~N. Page, ``Average entropy of a subsystem,''
  \href{http://dx.doi.org/10.1103/PhysRevLett.71.1291}{{\em Phys. Rev. Lett.}
  {\bfseries 71} (1993) 1291}.

\bibitem{Page93a}
D.~N. Page, ``Information in black hole radiation,''
  \href{http://dx.doi.org/10.1103/PhysRevLett.71.3743}{{\em Phys. Rev. Lett.}
  {\bfseries 71} (1993) 3743}.

\bibitem{Hayden07}
P.~Hayden and J.~Preskill, ``Black holes as mirrors: quantum information in
  random subsystems,''
  \href{http://dx.doi.org/10.1088/1126-6708/2007/09/120}{{\em JHEP} {\bfseries
  09} (2007) 120}.

\bibitem{Sekino08}
Y.~Sekino and L.~Susskind, ``Fast scramblers,''
  \href{http://dx.doi.org/10.1088/1126-6708/2008/10/065}{{\em JHEP} {\bfseries
  10} (2008) 065}.

\bibitem{Lashkari13}
N.~Lashkari, D.~Stanford, M.~Hastings, T.~Osborne, and P.~Hayden, ``Towards the
  fast scrambling conjecture,''
  \href{http://dx.doi.org/10.1007/JHEP04(2013)022}{{\em JHEP} {\bfseries 04}
  (2013) 22}.

\bibitem{Shenker:2013pqa}
S.~H. Shenker and D.~Stanford, ``{Black holes and the butterfly effect},''
  \href{http://dx.doi.org/10.1007/JHEP03(2014)067}{{\em JHEP} {\bfseries 03}
  (2014) 067}.

\bibitem{Maldacena:2016aa}
J.~Maldacena, S.~H. Shenker, and D.~Stanford, ``A bound on chaos,''
  \href{http://dx.doi.org/10.1007/JHEP08(2016)106}{{\em JHEP} {\bfseries 08}
  (2016) 106}.

\bibitem{Kitaev:2014t2}
A.~Kitaev, ``A simple model of quantum holography,'' 2015.
\newblock Talks at KITP, April 7, 2015 and May 27, 2015.

\bibitem{Harlow:2013tf}
D.~Harlow and P.~Hayden, ``{Quantum Computation vs. Firewalls},''
  \href{http://dx.doi.org/10.1007/JHEP06(2013)085}{{\em JHEP} {\bfseries 06}
  (2013) 085}.

\bibitem{Gao16}
P.~Gao, D.~L. Jafferis, and A.~Wall, ``Traversable wormholes via a double trace
  deformation.'' Arxiv:1608.05687.

\bibitem{Traversable2017}
J.~Maldacena, D.~Stanford, and Z.~Yang, ``Diving into traversable wormholes,''
  \href{http://dx.doi.org/10.1002/prop.201700034}{{\em Fortsch. Phys.}
  {\bfseries 65} (2017) 1700034}.

\bibitem{Roberts:2014isa}
D.~A. Roberts, D.~Stanford, and L.~Susskind, ``{Localized shocks},''
  \href{http://dx.doi.org/10.1007/JHEP03(2015)051}{{\em JHEP} {\bfseries 03}
  (2015) 051}.

\bibitem{Kitaev:2014t1}
A.~Kitaev, ``Hidden correlations in the hawking radiation and thermal noise,''
  2014.
\newblock Talk given at the Fundamental Physics Prize Symposium, Nov. 10, 2014.

\bibitem{Hosur:2015ylk}
P.~Hosur, X.-L. Qi, D.~A. Roberts, and B.~Yoshida, ``{Chaos in quantum
  channels},'' \href{http://dx.doi.org/10.1007/JHEP02(2016)004}{{\em JHEP}
  {\bfseries 02} (2016) 004}.

\bibitem{Roberts:2017aa}
D.~A. Roberts and B.~Yoshida, ``Chaos and complexity by design,''
  \href{http://dx.doi.org/10.1007/JHEP04(2017)121}{{\em JHEP} {\bfseries 04}
  (2017) 121}.

\bibitem{Grover:1996}
L.~K. Grover, \href{http://dx.doi.org/10.1145/237814.237866}{``A fast quantum
  mechanical algorithm for database search,''} in {\em Proceedings of the
  Twenty-eighth Annual ACM Symposium on Theory of Computing}, STOC '96,
  pp.~212--219.
\newblock ACM, New York, NY, USA, 1996.

\bibitem{Shenker:2013yza}
S.~H. Shenker and D.~Stanford, ``{Multiple Shocks},''
  \href{http://dx.doi.org/10.1007/JHEP12(2014)046}{{\em JHEP} {\bfseries 12}
  (2014) 046}.

\bibitem{BBBV}
C.~Bennett, E.~Bernstein, G.~Brassard, and U.~Vazirani, ``Strengths and
  weaknesses of quantum computing,''
  \href{http://dx.doi.org/10.1137/S0097539796300933}{{\em SIAM J. Comput.}
  {\bfseries 26} (2006) 1510}.

\end{mcitethebibliography}

\ifx\mcitethebibliography\mciteundefinedmacro
\PackageError{utphys.bst}{mciteplus.sty has not been loaded}
{This bibstyle requires the use of the mciteplus package.}\fi
\providecommand{\href}[2]{#2}\begingroup\raggedright\endgroup

\end{document}